\documentclass[twocolumn,english,superscriptaddress]{revtex4-2}
\usepackage[T1]{fontenc}
\usepackage{amsfonts}
\usepackage{amsmath}
\usepackage[latin9]{inputenc}
\usepackage{textcomp}
\usepackage{amstext}
\usepackage{subscript}
\usepackage{graphics}
\usepackage{graphicx}
\usepackage{float}
\usepackage{color}
\usepackage{pifont}
\usepackage{pmboxdraw}
\usepackage{amssymb}
\usepackage{esint}
\usepackage{tabularx}
\usepackage[export]{adjustbox}
\usepackage{hyperref}
\usepackage{upgreek,textgreek}
\usepackage{siunitx}
\usepackage{multirow}

\makeatletter

\newcommand{\lyxmathsym}[1]{\ifmmode\begingroup\def\b@ld{bold}
  \text{\ifx\math@version\b@ld\bfseries\fi#1}\endgroup\else#1\fi}

\usepackage{hyperref}
\usepackage{upgreek,textgreek}
\hypersetup{colorlinks=true,linkcolor=blue,citecolor=blue,urlcolor=blue}
\renewcommand{\fnum@figure}{FIG.~\thefigure}
\makeatother

\usepackage{babel}
\begin{document}
\title{Electrical and thermal transport properties of kagome metals \textit{A}Ti\textsubscript{3}Bi\textsubscript{5} (\textit{A} = Rb, Cs)}
\author{Xintong Chen}
\thanks {These authors contributed equally to this work.}
\affiliation{Low Temperature Physics Lab, College of Physics \& Center of Quantum
Materials and Devices, Chongqing University, Chongqing 401331, China}

\author{Xiangqi Liu \textcolor{blue}{\textsuperscript{*}}}
\affiliation{School of Physical Science and Technology, ShanghaiTech University, Shanghai 201210, China}

\author{Wei Xia}
\affiliation{School of Physical Science and Technology, ShanghaiTech University, Shanghai 201210, China}
\affiliation{
 ShanghaiTech Laboratory for Topological Physics, Shanghai 201210, China}

\author{Xinrun Mi}
\affiliation{Low Temperature Physics Lab, College of Physics \& Center of Quantum
Materials and Devices, Chongqing University, Chongqing 401331, China}

\author{Luyao Zhong}
\affiliation{Low Temperature Physics Lab, College of Physics \& Center of Quantum
Materials and Devices, Chongqing University, Chongqing 401331, China}

\author{Kunya Yang}
\affiliation{Low Temperature Physics Lab, College of Physics \& Center of Quantum
Materials and Devices, Chongqing University, Chongqing 401331, China}

\author{Long Zhang}
\affiliation{Low Temperature Physics Lab, College of Physics \& Center of Quantum
Materials and Devices, Chongqing University, Chongqing 401331, China}

\author{Yuhan Gan}
\affiliation{Low Temperature Physics Lab, College of Physics \& Center of Quantum
Materials and Devices, Chongqing University, Chongqing 401331, China}

\author{Yan Liu}
\affiliation{Analytical and Testing Center, Chongqing University, Chongqing 401331, China}

\author{Guiwen Wang}
\affiliation{Analytical and Testing Center, Chongqing University, Chongqing 401331, China}

\author{Aifeng Wang}
\affiliation{Low Temperature Physics Lab, College of Physics \& Center of Quantum
Materials and Devices, Chongqing University, Chongqing 401331, China}

\author{Yisheng Chai}
\affiliation{Low Temperature Physics Lab, College of Physics \& Center of Quantum
Materials and Devices, Chongqing University, Chongqing 401331, China}

\author{Junying Shen}
\affiliation{Institute of High Energy Physics, Chinese Academy of Sciences (CAS), Beijing 100049, China}
\affiliation{Spallation Neutron Source Science Center, Dongguan 523803, China}

\author{Xiaolong Yang}
\email{yangxl@cqu.edu.cn}
\affiliation{Low Temperature Physics Lab, College of Physics \& Center of Quantum
Materials and Devices, Chongqing University, Chongqing 401331, China}

\author{Yanfeng Guo}
\email{guoyf@shanghaitech.edu.cn}
\affiliation{School of Physical Science and Technology, ShanghaiTech University, Shanghai 201210, China}
\affiliation{
 ShanghaiTech Laboratory for Topological Physics, Shanghai 201210, China}
 
\author{Mingquan He}
\email{mingquan.he@cqu.edu.cn}
\affiliation{Low Temperature Physics Lab, College of Physics \& Center of Quantum
Materials and Devices, Chongqing University, Chongqing 401331, China}
\date{\today}

\begin{abstract}
We report electrical and thermal transport  properties of single crystalline kagome metals \textit{A}Ti\textsubscript{3}Bi\textsubscript{5} (\textit{A} = Rb, Cs).  Different from the structural similar kagome superconductors \textit{A}V$_3$Sb$_5$, no charge density wave instabilities are found in $A$Ti$_3$Bi$_5$.   At low temperatures below 5 K, signatures of superconductivity appear in $A$Ti$_3$Bi$_5$ as seen in magnetization measurements. However, bulk superconductivity is not evidenced by specific heat results. Similar to \textit{A}V$_3$Sb$_5$, $A$Ti$_3$Bi$_5$ show nonlinear magnetic field dependence of the Hall effect below about 70 K, pointing to a multiband nature. Unlike \textit{A}V$_3$Sb$_5$ in which phonons and electron-phonon coupling play important roles in thermal transport, the thermal conductivity in $A$Ti$_3$Bi$_5$ is dominated by electronic contributions. Moreover, our calculated electronic structure suggests that van Hove singularities are sitting well above the Fermi energy. Compared with \textit{A}V$_3$Sb$_5$, the absence of charge orders in $A$Ti$_3$Bi$_5$ is closely associated with minor contributions from electron-phonon coupling and/or van Hove singularities.

\end{abstract}

\maketitle
\section{Introduction}
The kagome lattice is a two-dimensional structure, which is made up of corner-sharing triangles \cite{Syozi1951}. Interestingly, the electronic structure of the kagome lattice hosts Dirac cones, van Hove singularities and a flat band \cite{Neupert2022}.  Various quantum phases of matter including quantum spin liquids, topological orders and unconventional superconductivity could arise from the kagome lattice by tuning the electron filling \cite{Brien2010,ko2009doped,Balents2010,kiesel2012sublattice,Kisel2013,Yu2012}. 
 Materials containing kagome lattice are thus prominent platforms to study the interplay of lattice, spin and charge degrees of freedom. Among various kagome materials, the recently discovered kagome metals $A$V$_3$Sb$_5$ ($A=$ K, Rb, Cs) are of particular interest \cite{Ortiz2019,Ortiz2020,Ortiz2021,Yin2021}. In $A$V$_3$Sb$_5$, V atoms form an ideal kagome structure at zoom temperature, giving rise to two van Hove singularities around the $M$ point, Dirac-like and $\mathbb{Z}_2$ topological bands near the Fermi level \cite{Li2021,Nakayama2021,Liu2021,Ortiz2021Cs,Hu_2022}. More interestingly,  a chiral charge density wave (CDW) associated with a giant anomalous Hall effect and time reversal symmetry breaking emerges below $T_\mathrm{CDW}\sim$ 80-100 K \cite{Jiang2021a,Wang2021CDW,Shumiya2021,Yang2020,Yu2021,Zheng2021_gating,Chendong2021,Zhou2021,Mielke2021,Yu2021b,Gan2021,Mi_2022,Liang2021CDW,Zhao2021a,Chen2021rotonpair,Xu2021,Li2021a,Hu2022co}. At low temperatures below $T_\mathrm{c}\sim$ 0.9-2.5 K, a superconducting phase arises and competes with the CDW phase \cite{Yu2021a,Yu2022pressure,Zhang_pressure,Du2021_pressure,Du2022_pressure,Zhu2021_pressure,Chen2021_pressure,ChenPRL_pressure,Wang2021_pressure,Oey_doping,Yang2021_doping}. Extensive studies are undergoing to explore the origins of the unusual charge orders and superconductivity.        

\begin{figure*}[t]
\centering
\includegraphics[scale=0.6]{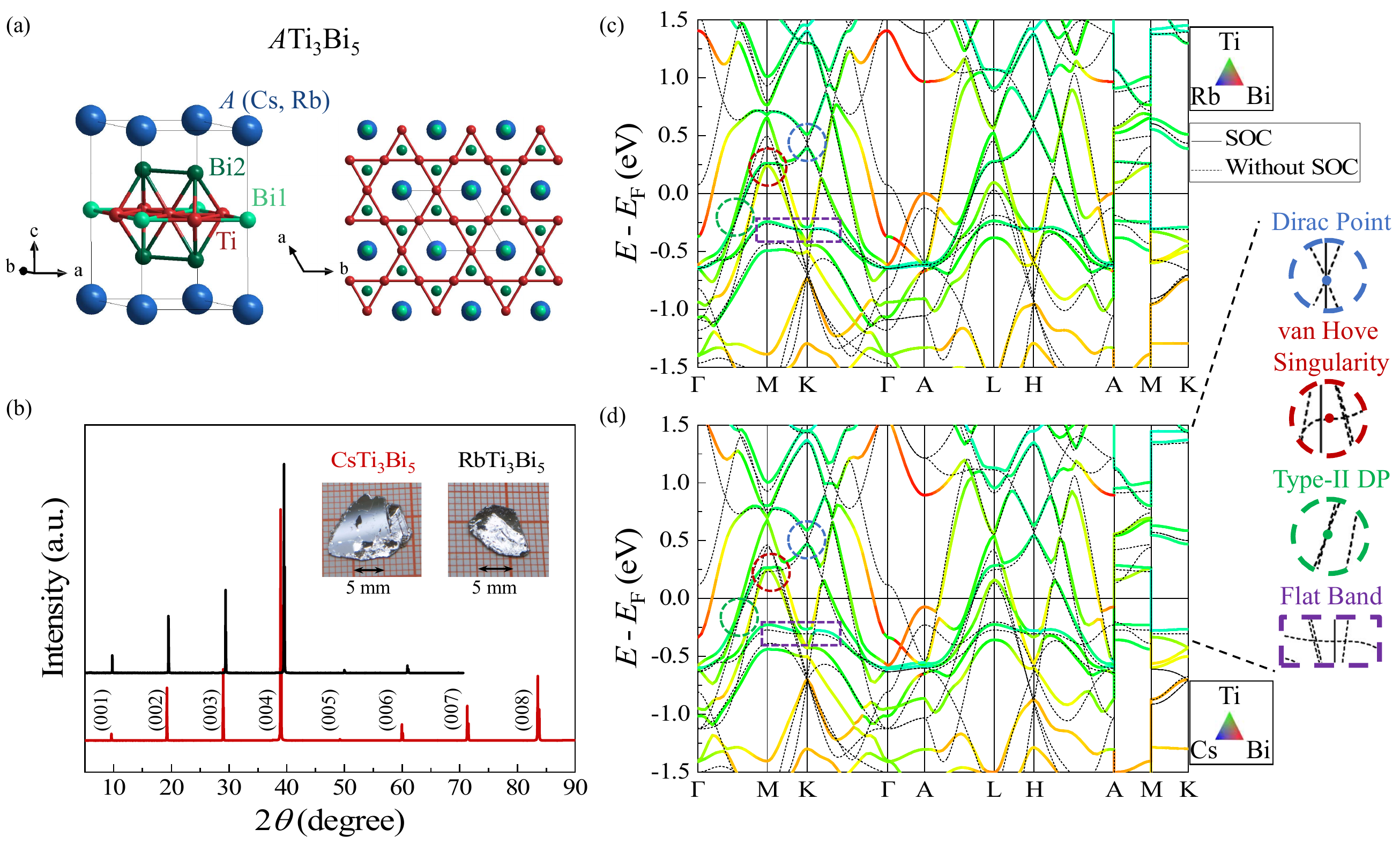}
\caption{ (a)  Side (left panel) and top (right panel) views of the crystal structure of $A$Ti$_3$Bi$_5$ (space group: $P6/mmm$). The Ti atoms form a kagome lattice in the Ti-Bi layer with one type of Bi atoms (Bi1) sitting in the center of kagome hexagons. The other type of Bi atoms (Bi2) locate above and below the Ti-Bi1 layer, forming a honeycomb pattern.  (b) X-ray diffraction patterns of RbTi$_3$Bi$_5$ (black curve) and CsTi$_3$Bi$_5$ (red curve) single crystals. The (00$L$) peaks can be nicely identified. Inset in (b) shows photographs of typical $A$Ti$_3$Bi$_5$ samples. (c) and (d) Calculated electronic band structures of RbTi$_3$Bi$_5$ and CsTi$_3$Bi$_5$ with (solid lines) and without (dash lines) considering spin-orbit coupling. }
\label{fig1}
\end{figure*}

The discoveries of kagome metals $A$V$_3$Sb$_5$  motivated much research effort to search for structural similar materials. Quite a few vanadium based kagome compounds, such as $A$V$_6$Sb$_6$, $A$V$_8$Sb$_{12}$, V$_6$Sb$_4$ and $R$V$_6$Sn$_6$, have been reported experimentally \cite{Shi2022_AVSb,Yin_2021VSb,Wang_2022VSb,Yang_2021VSb,Peng2021_RVSn,Husci}. 
In addition to theses systems, X.-W. Yi \textit{et al}. \cite{Yi_ATiBi} and Y. Jiang \textit{et al}. \cite{Jiang2022_ATiBi} predicted various thermodynamically stable $A$V$_3$Sb$_5$-like kagome materials based on first-principles calculations.  Notably, among the theoretically predicted $A$V$_3$Sb$_5$-like compounds, titanium-based kagome metals $A$Ti$_3$Bi$_5$ ($A$=Rb, Cs) have been experimentally synthesized lately \cite{Werhahn2022_ATiBi,Yang2022_ATiBi,Yang2022_ATiBi2,HuRbTiBi,Jiang2022_RbTiBi,Zhou2023RbTiBi}. Below 4 K, resistivity measurements on RbTi$_3$Bi$_5$ and CsTi$_3$Bi$_5$ revealed signatures of superconductivity, which were attributed to RbBi$_2$/CsBi$_2$ impurities by D. Werhahn \textit{et al}. \cite{Werhahn2022_ATiBi}. On the other hand, recent studies performed by H. Yang \textit{et al}. favor a bulk nature of superconductivity in CsTi$_3$Bi$_5$ \cite{Yang2022_ATiBi,Yang2022_ATiBi2}. Moreover, $\mathbb{Z}_2$ topological characters and electronic nematicity have been suggested in CsTi$_3$Bi$_5$ \cite{Yang2022_ATiBi,Yang2022_ATiBi2,HuRbTiBi}. Near the Fermi energy, flat bands and Dirac nodal lines have been identified in RbTi$_3$Bi$_5$ \cite{Jiang2022_RbTiBi,HuRbTiBi}.  \textcolor{black}{Using external stimulus, such as biaxial strain, van Hove singularities could be tuned to the Fermi level, which may lead to CDW and/or superconductivity \cite{Zhou2023RbTiBi}. It appears that $A$Ti$_3$Bi$_5$ is a prominent system to explore nematic order and topological properties. Importantly, $A$Ti$_3$Bi$_5$ does not show charge density wave instabilities \cite{Werhahn2022_ATiBi,Yang2022_ATiBi,Yang2022_ATiBi2,HuRbTiBi,Jiang2022_RbTiBi,Zhou2023RbTiBi}. Therefore,  comparing the similarities and differences between $A$Ti$_3$Bi$_5$ and $A$V$_3$Sb$_5$ can provide crucial clues for unraveling the mechanisms of CDW in $A$V$_3$Sb$_5$.} 

In this article, we present electrical and thermal transport measurements on $A$Ti$_3$Bi$_5$ ($A$=Rb, Cs) single crystals. No bulk superconductivity is found in our $A$Ti$_3$Bi$_5$ samples. Similar to $A$V$_3$Sb$_5$, the multiband electronic structure plays important roles in electrical transport of $A$Ti$_3$Bi$_5$. On the other hand, minor contributions from phonons and electron-phonon coupling are seen in thermal transport of $A$Ti$_3$Bi$_5$, which is different from that in $A$V$_3$Sb$_5$. One more difference between $A$Ti$_3$Bi$_5$ and  $A$V$_3$Sb$_5$ is that the separations between van Hove singularities and the Fermi level in $A$Ti$_3$Bi$_5$ are much larger than those in $A$V$_3$Sb$_5$. These differences may hold the key for the absence of CDW in $A$Ti$_3$Bi$_5$.

\section{Experimental Method}
Single crystals of $A$Ti$_3$Bi$_5$ samples were prepared by a self-flux method \cite{Jiang2022_RbTiBi}. Shiny crystals with in-plane sizes up to centimeters were obtained, as shown in the inset of Fig. \ref{fig1}(c).  Magnetization and heat capacity measurements were carried out in a Physical Property Measurement System (PPMS, Quantum Design Dynacool 9 T) using the vibrating sample magnetometer and relaxation method, respectively. The Seebeck effect and thermal conductivity were performed using the steady state method in PPMS on a home-made insert equipped with one heater and two-thermometer. The electrical transport measurements were recorded using the Hall bar geometry. The $A$Ti$_3$Bi$_5$ crystals degrade easily when exposed to ambient atmosphere. To prevent sample degradation, all experimental preparations were conducted in a glove box filled with argon.

Density functional theory (DFT) \cite{Kohn1965} calculations were performed to calculate the electronic band structure of CsTi$_3$Bi$_5$ and RbTi$_3$Bi$_5$ using the Vienna \textit{Ab initio} Simulation Package (VASP) \cite{Kresse1996,KRESSE199615}  with the projector augmented wave method (PAW) \cite{Bloch1994,Kresse1999}. The Perdew-Burke-Ernzerhof (PBE) parametrization of the generalized gradient approximation (GGA) was employed to treat the exchange-correlation energy functional \cite{Perdew1996}. \textcolor{black}{The plane wave cutoff energy of  520 eV was used, and a convergence threshold of $10^{-7}$ eV was adopted for each self-consistent electronic step. Cell parameters and internal atomic positions were fully relaxed with a $k$-mesh of $23\,{\times}\,23\,{\times}\,11$ until the maximum force on each atom was less than $10^{-3}$ eV {\AA}$^{-1}$. The optimized lattice constants are in good agreement with experiments \cite{Werhahn2022_ATiBi,Yang2022_ATiBi,Yang2022_ATiBi2}. To obtain the precise band structure and corresponding Fermi energy, a $23\,{\times}\,23\,{\times}\,11$ $\Gamma$-centered $k$-mesh was adopted based on an equilibrium structure.}

\begin{figure*}[t]
\centering  
\includegraphics[scale=0.6]{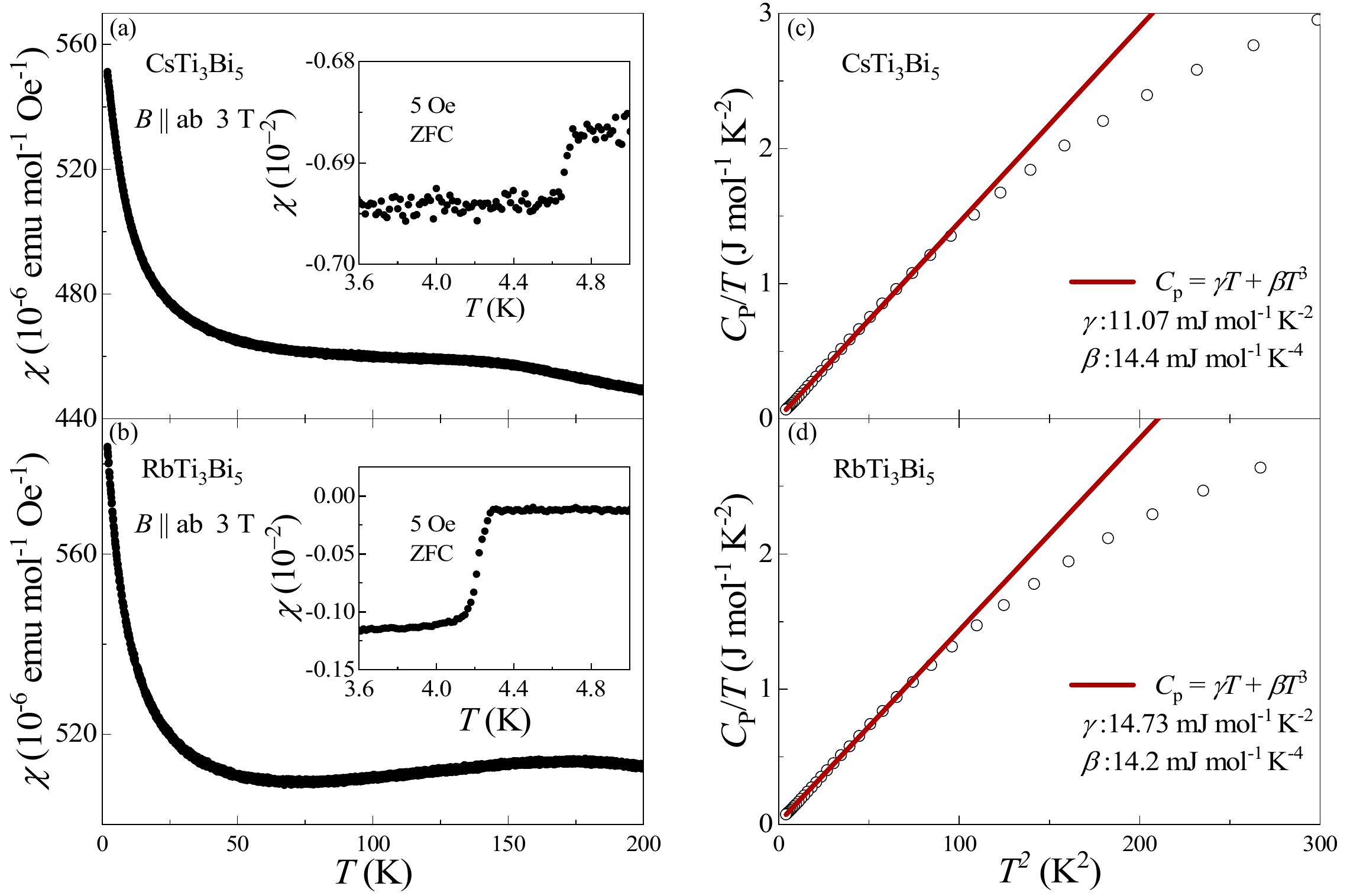}
\caption{(a) and (b) Temperature dependence of magnetization $M(T)$ for CsTi$_3$Bi$_5$ and RbTi$_3$Bi$_5$ samples. Insets in (a, b) show low-temperature zero-field-cooling (ZFC) measurements.   (c) and (d) Low-temperature Specific heat $C_\mathrm{p}$. The solid red lines in (c) and (d) are theoretical fitting in the form of $C_\mathrm{p}/T=\gamma+\beta T^2$.}
\label{fig2}
\end{figure*}

\section{Results and Discussion}

The crystal structure of $A$Ti$_3$Bi$_5$ is similar to that of $A$V$_3$Sb$_5$, as shown in Fig. \ref{fig1}(a). The Ti-Bi layers are sandwiched by the alkali $A$ layers, forming a layered structure within the $P6/mmm$ space group. The key element is the kagome net formed by Ti atoms, as seen in the right panel of Fig. \ref{fig1}(a). There are two types of Bi atoms with different coordination. The Bi1 atoms are located at the center of Ti kagome hexagons. The Bi2 atoms form a honeycomb pattern, which sits above and below the kagome lattice. The alkali atoms appear in a triangular form. In Fig. \ref{fig1}(b), the X-ray diffraction patterns of typical RbTi$_3$Bi$_5$ and CsTi$_3$Bi$_5$ single crystals are presented.  The (00$L$) peaks are clear seen in both crystals, suggesting the high quality of these samples. The lattice parameters of  RbTi$_3$Bi$_5$ and CsTi$_3$Bi$_5$ are found to be $a=5.8248$ {\AA}, $c=9.2498$ {\AA} and $a=5.8188$ {\AA}, $c=9.1507$ {\AA} respectively, agreeing well with earlier reports \cite{Werhahn2022_ATiBi,Yang2022_ATiBi,Yang2022_ATiBi2}.  \textcolor{black}{Figs. \ref{fig1}(c)(d) show the calculated electronic band structures of RbTi$_3$Bi$_5$ CsTi$_3$Bi$_5$. These two isostructural compounds basically show similar electronic band structures with slight differences in details.} Despite interruptions introduced by sizable spin-orbit coupling of the heavy Bi atoms, the featured Dirac points (DP), van Hove singularities (vHs) and flat bands of the kagome lattice are preserved. These  features have been experimentally identified in angle-resolved photoemission spectroscopy measurements \cite{Yang2022_ATiBi2,Jiang2022_RbTiBi,HuRbTiBi}. Similar to $A$V$_3$Sb$_5$, multiple bands mainly consisting of Ti 3$d$ and Bi $p_z$ orbitals cross the Fermi level. There are three electron pockets around the zone center $\Gamma$ point, two hole pockets near the zone boundary \cite{Jiang2022_RbTiBi,HuRbTiBi}. The multiband nature plays important roles in transport behavior [see Fig. \ref{fig:4}]. Compared with the V (3$d^3$) states in $A$V$_3$Sb$_5$, the Ti (3$d^2$) orbitals offer less electrons and lowers the Fermi energy ($E_F$). As a results, the van Hove singularities locating at the $M$ point are pushed well above $E_F$, while the flat bands come close to $E_F$. In $A$V$_3$Sb$_5$, it has been suggested that Fermi surface nesting promoted by van Hove singularities is responsible for the CDW instabilities \cite{Liu2021,Jiang2021a,Kang2022_FSN,Zhou2021_FSN,Hu2022}. The invisibility of van Hove singularities near $E_F$ may account for the absence of CDW in $A$Ti$_3$Bi$_5$.  

\begin{figure*}
\centering
\includegraphics[scale=0.6]{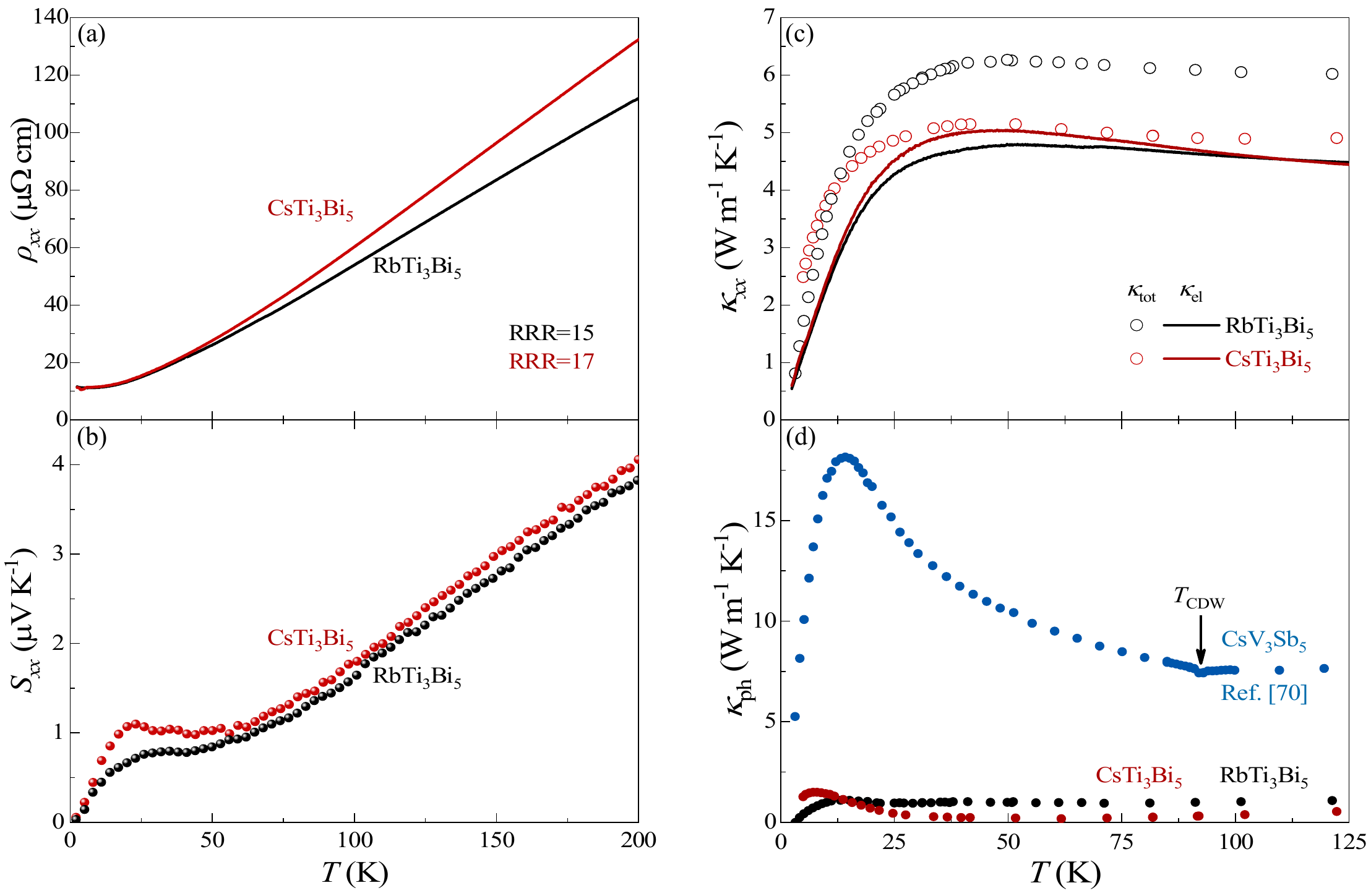}
\caption{(a-c) Temperature dependence of the longitudinal electrical resistivity ($\rho_{xx}$), Seebeck effect ($S_{xx}$)  and thermal conductivity ($\kappa_{xx}$) of CsTi$_3$Bi$_5$ and RbTi$_3$Bi$_5$ crystals. Circular points in (c) represent the total thermal conductivity ($\kappa_\mathrm{tot}$). Solid lines in (c) are calculated electronic thermal conductivity ($\kappa_\mathrm{el}$) according to the Wiedemann-Franz law. (d) Comparison of phonon thermal conductivity ($\kappa_\mathrm{ph}$) in  CsTi$_3$Bi$_5$, RbTi$_3$Bi$_5$ and CsV$_3$Sb$_5$ \cite{Yang2022AVSb}.}
\label{fig:3}
\end{figure*}

Figure \ref{fig2} presents the thermodynamic properties of CsTi$_3$Bi$_5$ and RbTi$_3$Bi$_5$. Unlike $A$V$_3$Sb$_5$, the $A$Ti$_3$Bi$_5$ family is rather robust against charge density wave instability. As shown in Figs. \ref{fig2}(a,b), the temperature-dependent magnetization of both compounds shows typical paramagnetic behaviors. No signatures of long-range magnetic orders or charge orders are seen, as also found in previous reports \cite{Werhahn2022_ATiBi,Yang2022_ATiBi,Yang2022_ATiBi2,Jiang2022_RbTiBi,Zhou2023RbTiBi}. At low temperatures, traces of Meissner effect are found below 4.8 and 4.3 K in  CsTi$_3$Bi$_5$ and RbTi$_3$Bi$_5$, respectively [see the insets in Figs. \ref{fig2}(a,b)], indicating possible emergence of superconductivity. The diamagnetic signal is, however, extremely weak. At 2 K, the superconducting volume fraction is less than 1\% in an external magnetic field of 5 Oe. It is likely that the observed weak Meissner signals are originated from CsBi$_2$ and RbBi$_2$ impurities, which become superconducting below 4.65 and 4.21 K, respectively \cite{Werhahn2022_ATiBi,Csbi2}. As seen in Figs. \ref{fig2}(e,f), no evidence of superconducting transition can be identified in specific heat. This further suggests that bulk superconductivity is absent in samples studied here.  By analyzing the low-temperature specific heat according to $C_\mathrm{p}=\gamma T+\beta T^3$, the electronic specific heat (Sommerfeld) coefficients $\gamma$ are estimated to be 11.1(2) and 14.7(3) mJ mol$^{-1}$ K$^{-2}$ for CsTi$_3$Bi$_5$ and RbTi$_3$Bi$_5$, respectively. The phonon specific heat coefficients $\beta$ read 14.4(3) and 14.2(3) mJ mol$^{-1}$ K$^{-4}$ for CsTi$_3$Bi$_5$ and RbTi$_3$Bi$_5$, respectively. The Debye temperature $\Theta_\mathrm{D}$ can be evaluated accordingly following $\beta=12\pi^4NR/5\Theta_\mathrm{D}^3$ with $N$ and $R$ being the number of atoms per unit cell and the ideal gas constant. It is found that $\Theta_\mathrm{D}$=106.63 and 107.13 K for CsTi$_3$Bi$_5$ and RbTi$_3$Bi$_5$, respectively.

The in-plane resistivity ($\rho_{xx}$) of CsTi$_3$Bi$_5$ and RbTi$_3$Bi$_5$ samples measured in zero field is presented in Fig. \ref{fig:3}(a). Both compounds show metallic behaviors. The residual resistivity ratios RRR=$\rho_{xx}$(300 K)/$\rho_{xx}$(2 K) are 17 and 15 for CsTi$_3$Bi$_5$ and RbTi$_3$Bi$_5$ samples, respectively.  The thermoelectric Seebeck ($S_{xx}$) properties are shown in Fig. \ref{fig:3}(b). Positive values of $S_{xx}$ are found all the way from room temperature down to 2 K in both samples, implying the dominant roles played by hole-like carriers. In $A$V$_3$Sb$_5$, on the other hand, electron-like and hole-like excitations compete with each other, leading to sign changes in the Seebeck signal and Hall coefficient at low temperatures \cite{Gan2021,Mi_2022}. Compared with $A$V$_3$Sb$_5$, the Fermi levels of $A$Ti$_3$Bi$_5$ shift downwards significantly [see Fig. \ref{fig1}(d)]. As a result, the electron pockets centering at the zone center are reduced, while the hole pockets near the zone boundaries are enlarged. Therefore, it is not unexpected that hole-like carries play major roles in $A$Ti$_3$Bi$_5$. 

The thermal transport properties of $A$Ti$_3$Bi$_5$ are also rather different from those in $A$V$_3$Sb$_5$. As displayed in Fig. \ref{fig:3}(c), the total longitudinal thermal conductivity $\kappa_\mathrm{tot}$ of CsTi$_3$Bi$_5$ and RbTi$_3$Bi$_5$ is dominated by electronic contributions ($\kappa_\mathrm{el}$). The electronic thermal conductivity is estimated from the Wiedemann-Franz law via $\kappa_\mathrm{el}(T)=\sigma L_0 T$, where $\sigma$ and $L_0$ are electronic conductivity and the Lorenz number. Fig. \ref{fig:3}(d) compares the phononic thermal conductivity $\kappa_\mathrm{ph}=\kappa_\mathrm{tot}-\kappa_\mathrm{el}$ of CsTi$_3$Bi$_5$, RbTi$_3$Bi$_5$ and CsV$_3$Sb$_5$. Clearly, $\kappa_\mathrm{ph}$ of CsV$_3$Sb$_5$ is one order of magnitude larger than that in $A$Ti$_3$Bi$_5$. In the charge ordered state below $T_\mathrm{CDW}$, $\kappa_\mathrm{ph}$ of CsV$_3$Sb$_5$ shows typical behaviours of phononic heat transport. Above $T_\mathrm{CDW}$, sizable charge fluctuations and electron-phonon coupling lead to glass-like thermal conductivity which increases linearly with warming \cite{Yang2022AVSb}. In $A$Ti$_3$Bi$_5$, $\kappa_\mathrm{ph}$ depends weakly on temperature and $\kappa_\mathrm{el}$ dominates in heat conduction. The subdominant roles played by phonons and electron-phonon coupling in $A$Ti$_3$Bi$_5$ may also represent important factors for the absence of charge orders.   

\begin{figure*}
\centering
\includegraphics[scale=0.6]{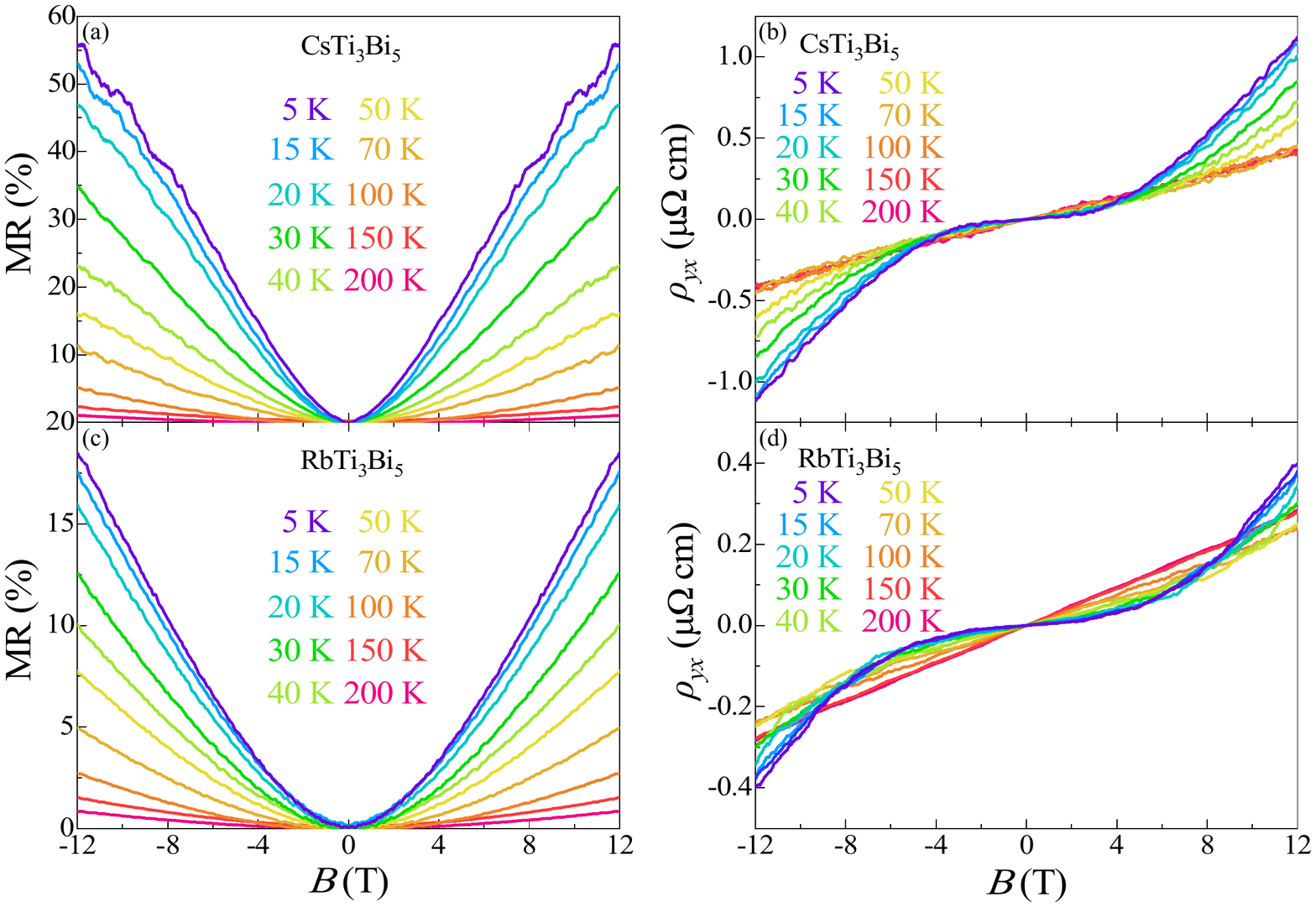}
\caption{(a, c) and (b, d) Magnetoresistance (MR) and Hall resistivity ($\rho_{yx}$) of CsTi$_3$Bi$_5$ and RbTi$_3$Bi$_5$ recorded at selective temperatures. Nonlinear $\rho_{yx}(B)$ curves appear below 70 K in both materials. }
\label{fig:4}
\end{figure*}
\begin{figure*}
\centering
\includegraphics[scale=0.6]{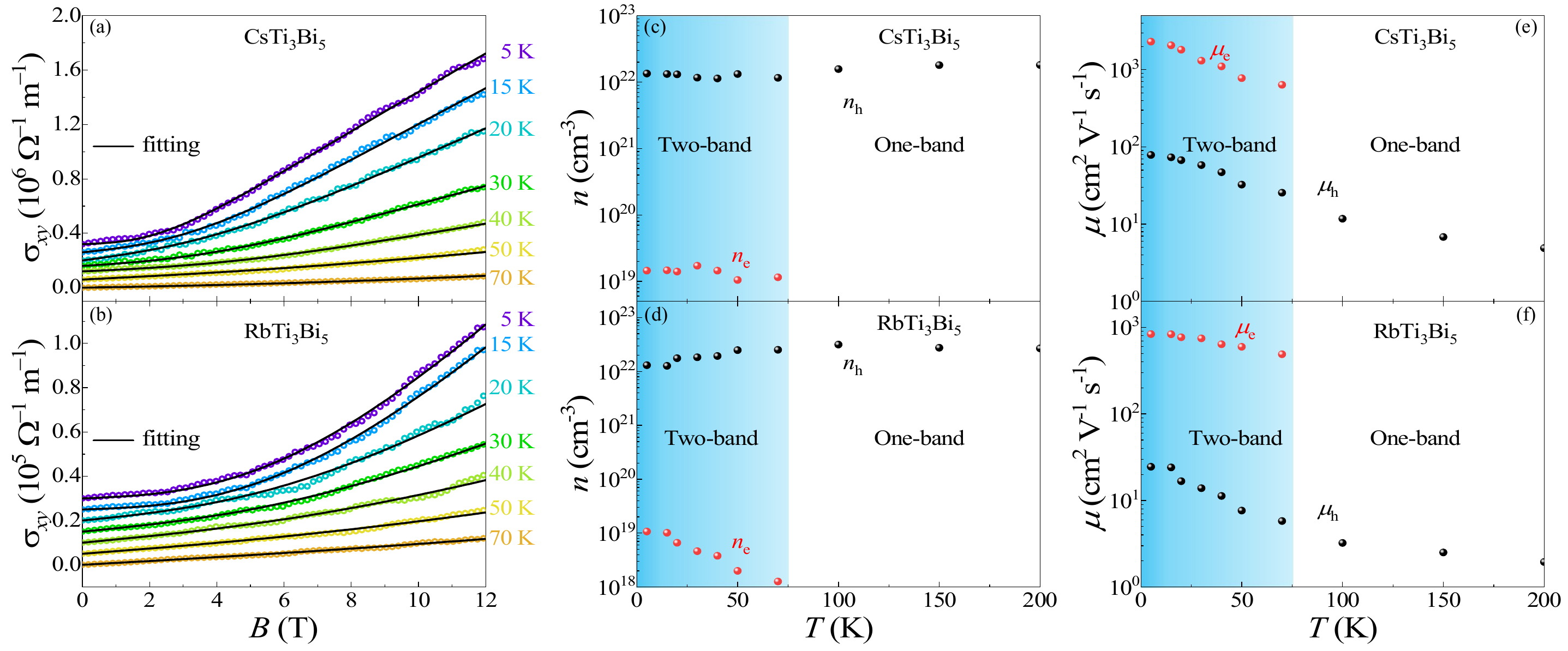}
\caption{(a) and (b) Electrical Hall conductivity ($\sigma_{xy}$) of  CsTi$_3$Bi$_5$ and RbTi$_3$Bi$_5$. Scattered circular points are experimental data. Solid lines are theoretical fittings using a two-band model. Vertical offsets have been applied for clarity. (c-f) Temperature dependence of carrier density and mobility of each band obtained from the Hall conductivity. } 
\label{fig:5}
\end{figure*}

In Fig. \ref{fig:4}, we present isothermal magnetoresistance (MR) and Hall resistivity ($\rho_{yx}$) of $A$Ti$_3$Bi$_5$. In both materials, positive MR starts to develop below 200 K and becomes more apparent at low temperatures. Large MR$=[\rho_{xx}(B)-\rho_{xx}(0)]/\rho_{xx}(0)$ reaching 55\% in 12 T at 5 K is found in CsTi$_3$Bi$_5$. In RbTi$_3$Bi$_5$, lower MR values less than 20\% is observed at the same conditions. Weak signatures of quantum oscillations are also seen in CsTi$_3$Bi$_5$ at low temperatures and in high magnetic fields. In both materials above 70 K, the Hall resistivity depends linearly on magnetic field with positive slops. This implies that the electrical transport is dominated by a single hole band at high temperatures. Further cooling below 70 K, $\rho_{yx}(B)$ becomes nonlinear, pointing to multiband transport. Unlike $A$V$_3$Sb$_5$, no sign changes are found in the temperature-dependent Hall coefficient of $A$Ti$_3$Bi$_5$, in accordance with the Seebeck results. Notably, below 20 K, $\rho_{yx}(B)$ shows an 'S'-shaped appearance, which is reminiscent to the anomalous Hall effect found in the CDW phase of $A$V$_3$Sb$_5$ \cite{Yang2020,Yu2021}. The nontrivial electronic structures near $E_F$, such as Dirac nodal lines, may produce  an anomalous Hall effect.   On the other hand, multiband transport can also give rise to such curvatures in $\rho_{yx}(B)$. Considering the multiband nature of $A$Ti$_3$Bi$_5$, we use a two-band picture to describe the Hall conductivity

\begin{equation}
    \sigma_{xy}(B)=\frac{n_ee\mu_e^2B}{1+\mu_e^2B^2}+\frac{n_he\mu_h^2B}{1+\mu_h^2B^2},
\end{equation}
where $n_{e(h)}$, $\mu_{e(h)}$ are carrier density and mobility of the corresponding electron (hole) pocket. The experimental Hall conductivity was evaluated from the resistivity data using $\sigma_{xy}=-\rho_{yx}/(\rho_{xx}^2+\rho_{yx}^2)$. \textcolor{black}{In the two-band fitting process, the constraint for longitudinal electrical conductivity ($\sigma_{xx}$) in zero magnetic field was also applied, i.e., $\sigma_{xx}(B=0)=n_ee\mu_e+n_he\mu_h$.} As displayed in Figs. \ref{fig:5}(a, b), the two-band approximation well describes the sublinear Hall conductivity $\sigma_{xy}(B)$ curves below 70 K. It is very likely that the nonlinear magnetic field dependence of the Hall effect observed in $A$Ti$_3$Bi$_5$ is originated from multiband transport.  From the two-band analysis, we obtain the temperature-dependent carrier density and mobility of each band, as shown in Figs. \ref{fig:5}(c-f). In both compounds above 70 K, a single hole band dominates in transport with weakly temperature-dependent carrier density and mobility. Below 70 K, an electron band with lower concentration but higher mobility comes into play. \textcolor{black}{Still, holelike carriers dominate at low temperatures, in agreement with the positive Seebeck signal seen all the way from room temperature to 2 K [see Fig. \ref{fig:3}(b)]. As shown in Figs. \ref{fig1}(c,d), the hole pockets around the zone boundary are mainly contributed from Bi $d$-orbitals. Therefore, the transport properties of $A$Ti$_3$Bi$_5$ are dominated by Bi $d$-orbitals. Similarly in $A$V$_3$Sb$_5$, the V $d$-orbitals play dominant roles in transport properties and the formation of CDW \cite{Ortiz2021Cs,Jiang_kagome}. In $A$V$_3$Sb$_5$, multiband transport effects also appear at low temperatures below about 50 K \cite{Gan2021,Mi_2022}. The multiband electronic structures thus play important roles in transport behaviors of $A$Ti$_3$Bi$_5$ and $A$V$_3$Sb$_5$, despite their different ground states. These results suggest that, in non-magnetic multiband metals, 'S'-shaped non-linear Hall response may not necessarily originate from anomalous Hall effect, and contributions from multiband transport can not be neglected. }

\section{Conclusions}
We have studied the electrical and thermal transport behaviours of the kagome metals $A$Ti$_3$Bi$_5$. The structural similar $A$Ti$_3$Bi$_5$ and $A$V$_3$Sb$_5$ families share few similarities but differ significantly. Both $A$Ti$_3$Bi$_5$ and $A$V$_3$Sb$_5$ systems host multiband electronic structures, which are manifested in the nonlinear magnetic-field-dependent Hall effect at low temperatures. Unlike $A$V$_3$Sb$_5$,  van Hove singularities in $A$Ti$_3$Bi$_5$ are away from the Fermi level. In addition, the heat conduction in $A$Ti$_3$Bi$_5$ is mainly carried by charge carriers, which is different from $A$V$_3$Sb$_5$ in which phonons and electron-phonon coupling contribute significantly. These differences between $A$Ti$_3$Bi$_5$ and $A$V$_3$Sb$_5$ can provide important insights on the driving force of CDW in $A$V$_3$Sb$_5$. 

\section{Acknowledgments}
 This work has been supported by National Natural Science Foundation of China (Grant Nos. 11904040, 52125103, 52071041, 12004254, 12004056, 11674384, 11974065), Chinesisch-Deutsche Mobilit\"atsprogamm of Chinesisch-Deutsche Zentrum f\"ur Wissenschaftsf\"orderung (Grant No. M-0496), Chongqing Research Program of Basic Research and Frontier Technology, China (Grant No. cstc2020jcyj-msxmX0263). Y. Guo acknowledges the support by the Major Research Plan of the National Natural Science Foundation
of China (No. 92065201).

\bibliographystyle{apsrev4-2}

\end{document}